\documentclass[pre,aps,twocolumn]{revtex4-1}
\usepackage{graphicx}
\usepackage{bm}
\usepackage{tabularx}
\usepackage{threeparttable}
\usepackage{amssymb}

\begin{document}

\title{On the nature of highly vibrationally excited states of Thiophosgene}


\author{Srihari Keshavamurthy}
\affiliation{Department of Chemistry, Indian Institute
of Technology, Kanpur, Uttar Pradesh 208016, India}



\begin{abstract}
In this work an analysis of the highly vibrationally excited states of thiophosgene (SCCl$_{2}$) is made in order to gain insights into some of the experimental observations and spectral features. The states analyzed herein lie in a spectrally complex region where strong mode mixings are expected due to the overlap of several strong anharmonic Fermi resonances.  Two recent techniques, a semiclassical angle space representation of the eigenstates and the parametric variation of the eigenvalues (level-velocities) are used to identify eigenstate sequences exhibiting common localization characteristics. Preliminary results on the influence of highly excited out-of-plane bending modes on the nature of the eigenstates suggest a possible bifurcation in the system.
\end{abstract}

\maketitle
 
\section{Introducion}

Understanding the nature of the highly excited molecular eigenstates is equivalent to deciphering the mechanism of intramolecular vibrational energy redistribution (IVR) occuring in the molecule\cite{smalley83}. However, the assignment of eigenstates is far from simple. The existence of and interplay of several strong anharmonic resonances  result in complicated spectral patterns and highly mixed states. Nevertheless, eigenstates of molecular systems rarely exhibit the extreme scenario of being completely ergodic or completely regular\cite{bnn04}. A generic situation, even at fairly high energies, is the coexistence of several classes of eigenstates with differing degree of mixing being interspersed among each other\cite{jajcp99,gl08}. More interestingly, the interspersed states can show several sequences associated with certain identifiable common localization characteristics. In other words, although a full set of quantum numbers do not exist for labeling such states, there are reasons to believe that a sufficient number of approximate quantum numbers do exist to understand, and hence assign, the eigenstates. Since the approximate quantum numbers arise out of the local dynamics due to specific resonances, relevant at the energies of interest, the assignment is inherently dynamical in nature. In a nutshell, several decades of work have shown that dynamical assignments can be done reliably only if the structure of the corresponding classical phase space is well understood\cite{dynassign}.

This work is concerned with the dynamical assignment of the eigenstates of  thiophosgene SCCl$_{2}$.  The Hamiltonian of interest is a highly accurate spectroscopic Hamiltonian obtained\cite{sgjcp06} by Gruebele and Sibert via a canonical Van Vleck perturbation analysis of the experimentally derived normal mode potential surface\cite{sgpccp04}. The Hamiltonian can be expressed as $H=H_{0}+V_{\rm res}$ with

\begin{eqnarray}
H_{0} & = &\sum_{i} \omega_{i} \left(v_{i} + \frac{1}{2} \right) + 
             \sum_{ij} x_{ij} \left(v_{i} + \frac{1}{2} \right) \left(v_{j} + \frac{1}{2} \right) \nonumber \\
        &+& \sum_{ijk} x_{ijk} \left(v_{i} + \frac{1}{2} \right) \left(v_{j} + \frac{1}{2} \right) \left(v_{k} + \frac{1}{2} \right),  \nonumber \\
V_{\rm res} &=& K_{526} a_{2}^{\dagger} a_{5} a_{6}^{\dagger} + K_{156} a_{1} a_{5}^{\dagger} a_{6}^{\dagger} \nonumber \\
                   &+& K_{125} a_{1}^{\dagger} a_{2}^{\dagger} a_{5}^{2} + K_{36} a_{3}^{2} a_{6}^{\dagger 2} \nonumber \\
                  &+ & K_{231} a_{1}^{\dagger} a_{2} a_{3}^{2} + K_{261} a_{1}^{\dagger} a_{2} a_{6}^{2} + {\rm c.c}.
\label{qham}
\end{eqnarray}

The zeroth-order anharmonic Hamiltonian is $H_{0}$, whose eigenstates are the feature states and $V_{\rm res}$ accounts for the important anharmonic resonances responsible for the IVR dynamics. In the rest of the paper the various resonances will be refered to by simply indicating the modes. For instance, the first resonance term in $V_{\rm res}$ above will be refered to as the $526$-resonance.
The hermitian adjoints of the operators are denoted by c.c., and the operators $a_{k}$, $a_{k}^{\dagger}$, and $v_{k} \equiv a_{k}^{\dagger} a_{k}$ are the
lowering, raising, and number operators for the $k^{\rm th}$ mode respectively with $[a_{k},a_{k}^{\dagger}]=1$, in analogy to the
usual harmonic oscillator operators. Based on the possible resonances suggested by the normal mode frequencies 
and the observed spectral ``clumps" in the dispersed fluorescence spectra, three approximately conserved quantities (known as
polyads) arise which can be expressed as
\begin{eqnarray}
K &=& v_{1} + v_{2} + v_{3} , \nonumber \\
L &=& 2v_{1} + v_{3} + v_{5} + v_{6},  \nonumber \\
M &=& v_{4}.
\label{polyads}
\end{eqnarray}
Thus, the $\nu_{4}$ mode
is not involved in any resonant interactions with the rest of the modes.  Due to the
polyad constraints one has effectively a three degree of freedom system and three of the quantum numbers out of a total six needed to 
assign a quantum state are already known.  
Nevertheless, as will be seen below, even this reduced system
presents a stiff challenge with regards to the dynamical assignments of the highly excited eigenstates.

The IVR dynamics in SCCl$_{2}$  has been the subject of several recent experimental\cite{sgpccp04,cgjcp109,rmjcp08} and theoretical\cite{jtsjpc06,cgjcp209,jtjcp10} studies.
In an early detailed study, Strickler and Gruebele established\cite{sgpccp04} that the IVR state space (zeroth-order quantum number space) has an effective dimensionality of three as opposed to the theoretically possible six.  Combined with the observation of regular progressions persisting upto the first dissociation energy of about $600$ THz, one expects restricted IVR in the system. This in itself is
surprising since the effective Hamiltonian fit, as in eq.~\ref{qham}, to the stimulated emission pumping (SEP) spectra contains strong multimode Fermi resonances. In addition, the
SEP spectra are quite complex and congested in the $200-300$ THz region. However, it was noted that the regular progressions
mainly involved the Franck-Condon active $\nu_{1}$ (C-S stretch) and the $\nu_{4}$ (out-of-plane bend) modes,
and thus avoided the strongest resonances. Furthermore, 
Chowdary and Gruebele recently\cite{cgjcp109} showed the existence of sharp SEP spectral features in the energy range of $550-700$ THz which includes both the
molecular SCCl$_{2}$ $\rightarrow$ CS $+$ Cl$_{2}$ and the radical SCCl$_{2}$ $\rightarrow$ SCCl $+$ Cl dissociation channels.  Although, one could rationalize the observation as due to the avoidance of the strong
$526$-resonance involving modes $\nu_{2}$ ($a_{1}$ symmetry C-Cl stretch), $\nu_{5}$ ($b_{2}$ symmetry C-Cl stretch), and
$\nu_{6}$ ($b_{2}$ symmetry Cl-C-Cl bend), a similar argument for the equally strong $156$-resonance
involving modes $\nu_{1}$ ($a_{1}$ symmetry C=S stretch), $\nu_{5}$, and $\nu_{6}$ is not possible. 
Sibert and Gruebele
performed\cite{sgjcp06} a detailed study of the IVR dynamics using eq.~\ref{qham} and concluded that around $240$ THz ($\sim$ $8000$ cm$^{-1}$),
close to the threshold for facile IVR,
zeroth-order states coexist exhibiting a broad range of dynamical behvior ranging from fairly restricted to highly facile IVR. A rather
extensive study\cite{cgjcp209} of nearly $10^{6}$ zeroth-order bright (``feature") states 
 by Chowdary and Gruebele concluded that about one
in thousand feature states are localized even at the highest energy - not enough to invalidate statistical theories but certainly indicating
that even near the dissociation limit SCCl$_{2}$ is not just a bag-of-atoms.

The above observations  hint at the existence of different classes of eigenstates in a specific energy range. For instance, even around an energy of about $590$ THz one observes\cite{cgjcp109} states whose IVR line
widths vary nonmonotonically with energy. Thus, in order to gain further insights into the IVR dynamics a 
closer look at the eigenstates of eq.~\ref{qham} is necessary. An earlier work by Jung, Taylor, and Sibert using
a second order version of eq.~\ref{qham} established\cite{jtsjpc06} that several distinct eigenstate classes exist
in the spectrally complicated region $7000-9000$ cm$^{-1}$ ($\sim$ $210-270$ THz).   More recently\cite{jtjcp10}, Jung and Taylor dynamically assigned the states and came up with a detailed semiclassical classification scheme for the
eigenstates. 

In this work an attempt is made to answer the following questions. Firstly, can one identify the various eigenstate classes in a manner that
is not dependent on visualizing the high dimensional wavefunctions quantum mechnically or semiclassically? More importantly, such a 
method should be able to make contact with the relevant structures like periodic orbits and their bifurcations in the
high dimensional classical phase space. Secondly, do the eigenstate classes and sequences identified previously persist when using a higher order and more accurate Hamiltonian?   Finally, is there a possibility of new modes being born due to specific bifurcations in the
system? 

\subsection {Classical limit Hamiltonian}

In order to dynamically assign the eigenstates it is important to analyze the classical limit Hamiltonian corresponding to eq.~\ref{qham} and the procedure is well established in literature. The full classical Hamiltonian, in terms of the zeroth-order action-angle variables $({\bf I},{\bm \theta})$, can be found in an earlier work\cite{jtsjpc06}. Here, the reduced three degree of freedom
classical Hamiltonian, taking into cognizance the polyads $(K,L,M)$, is of interest and can be obtained via a suitable $({\bf I},{\bm \theta}) \rightarrow ({\bf J},{\bm \psi})$ canonical transformation\cite{jtsjpc06}. The resulting reduced Hamiltonian can be expressed as $H({\bf J},{\bm \psi})=H_{0}({\bf J})+V_{\rm res}({\bf J},{\bm \psi})$ where
\begin{eqnarray}
H_{0}({\bf J},{\bm \psi}) &=& C + \sum_{i=1}^{3} \varpi_{i} J_{i} + \sum_{i,j = 1}^{3} \alpha_{ij} J_{i}J_{j} \nonumber \\
                                      &+& \sum_{i,j,k = 1}^{3} \beta_{ijk}  J_{i} J_{j} J_{k},
\end{eqnarray}
is the zeroth-order Hamiltonian and the resonant perturbations (nonlinear resonances) are given by
\begin{eqnarray}
V_{\rm res}({\bf J},{\bm \psi}) &=& v_{156}({\bf J};K_{c},L_{c}) \cos \psi_{1} \nonumber \\
                                              &+& v_{526}({\bf J};K_{c},L_{c}) \cos \psi_{2} \nonumber \\
                                              &+& v_{125}({\bf J};K_{c})\cos(\psi_{1}+\psi_{2}) \nonumber \\
                                              &+& v_{36}({\bf J};K_{c},L_{c})\cos 2\psi_{3} \nonumber \\
                                              &+& v_{231}({\bf J}) \cos(\psi_{1}-\psi_{2}-2\psi_{3}) \nonumber \\
                                              &+& v_{261}({\bf J};K_{c},L_{c})\cos(\psi_{1}-\psi_{2}).
\end{eqnarray}
In the above equations the actions conjugate to the angles $(\psi_{4},\psi_{5},\psi_{6})$
are $(K_{c}=K+3/2,L_{c}=L+5/2,M_{c}=M+1/2)$ {\it i.e.,} the classical analogs of the quantum polyads. The transformed actions $\{J_{k}\}$ are related to the zeroth-order quantum numbers as $J_{k} = (v_{k}+1/2)\hbar$.
The various parameters $(C,{\bm \varpi},{\bm \alpha},{\bm \beta})$ of the reduced Hamiltonian are
determined in terms of the original parameters (see supplementary information in reference \onlinecite{sgjcp06} for the values). The relations
are not shown here but note that at this order
the parameters $(C,{\bm \varpi},{\bm \alpha})$ depend on the polyads $K,L$, and $M$.

\section{Dynamical assignment}

In this section two recent techniques\cite{jtjpc07,sejcp03}, used to analyze the highly excited eigenstates, are briefly mentioned. 
The reader is referred to an earlier publication\cite{mkjpc09} for the theoretical basis and validity of the methods along with a detailed comparison.

In the first approach\cite{jtsjpc06,jtjcp10,jtjpc07} the quantum states $| \alpha \rangle$ for a given $(K,L,M)$ are expressed in the semiclassical angle space ${\bm \psi}=(\psi_{1},\psi_{2},\psi_{3})$ as
\begin{equation}
\langle {\bm \psi} | \alpha \rangle = \sum_{{\bf v} \in (K,L,M)} C_{\alpha {\bf v}} e^{i {\bf v} \cdot {\bm \psi}},
\label{semirep}
\end{equation}
with ${\bf v}=(v_{1},v_{2},v_{3})$ and ignoring an overall phase factor. Due to the $2\pi$-periodicity of the angles, the semiclassical wavefunctions $\alpha({\bm \psi})$ now ``live" on the classical reduced configuration space which is a three  dimensional torus $T^{3}$. The nature of the states is determined by plotting the density $|\langle {\bm \psi}|\alpha \rangle|^{2}$ and the phase of eq.~\ref{semirep} on select two-dimensional sections of $T^{3}$. 

The second approach\cite{sejcp03,mkjpc09} involves computing the eigenstate expectation values of the various anharmonic resonances.  The expectation values are naturally related to the parametric variation of the corresponding eigenvalues, also called as the level-velocities. For example, 
\begin{equation}
\langle \alpha | \hat{V}_{156} | \alpha \rangle = \frac{\partial E_{\alpha}}{\partial K_{156}}.
\end{equation}
Previous works\cite{sejcp03,skcpl04,mkjpc09,peres} have shown that the level-velocities, semiclassically, are extremely sensitive to the underlying structures in the classical phase space. There is also an interesting connection (not elaborated here) between the parametric variation of eigenvalues to the matrix fluctuation-dissipation theorem\cite{gjpc96} of Gruebele.
In what follows the terms expectation values and level-velocities will be used interchangeably. 

Recently\cite{mkjpc09} it was argued that a combination of the above approaches can be very powerful towards gaining key insights into the nature of the highly excited eigenstates. The current work provides further support for such an argument. To begin with, a single resonance case is analyzed in order to set the stage for looking at select eigenstates of the full system. Although the single resonance case is classically integrable, the correct assignment of the states needs to be done with care due to the multimode nature of the resonance. 

\subsection{Integrable multimode resonance}

In the full quantum Hamiltonian, all the anharmonic resonances as shown in in eq.~\ref{qham} are turned off except for the multimode $156$-resonance. Since, the quantum and classical Hamiltonians couple only the $\nu_{1},\nu_{5}$, and $\nu_{6}$ modes, the quantum numbers $(v_{2},v_{3})$ are good quantum numbers. Consequently, one already has five good quantum numbers $(K,L,M,v_{2},v_{3})$ to label those quantum states that are under the influence of the said resonance. States that are not affected by the resonance are trivially assigned by the full set of zeroth-order quantum numbers and hence will not be discussed further.  In the present case it is desirable to organize the states using $P_{s} \equiv (L-v_{3})$ since this represents the plane on which the states live in the original six-dimensional state space. 

\begin{figure}[h]
\includegraphics[height=3.0in,width=4.0in]{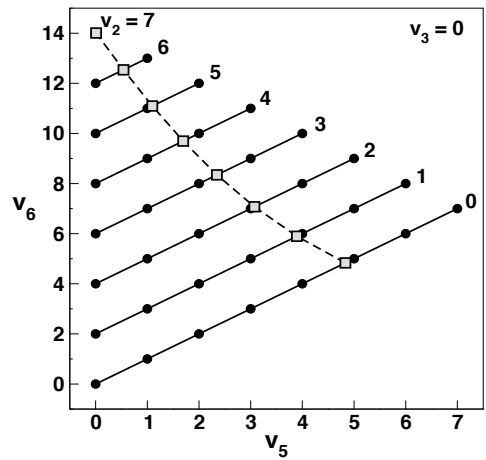}
\caption{Projection of the $P_{s}=14$ plane onto the $(v_{5},v_{6})$ space.  Mixing of the zeroth-order states occurs only along lines corresponding to constant values of $v_{2}$ (indicated)  leading to the sequences observed in Fig.~\ref{lvelint156plot}(A) and (B).  Classical prediction (cf. Eq.~\ref{clvel156predict}, grey squares) of the maximal level-velocities corresponding to large quantum eigenstate expectation value $\langle \hat{V}_{156} \rangle$ of the $156$-resonance operator.}
\label{stspgeomplot}
\end{figure}

The geometry of the $(v_{1},v_{5},v_{6})$ space is shown in Fig.~\ref{stspgeomplot} for the specific case of $v_{3}=0$. For given $(K,L,M)$, the space is filled with $P_{s}$-planes with $v_{3}$ being fixed on each plane. On a given $P_{s}$-plane, fixed values of $v_{2}$ correspond to lines determined by points $(v_{1},K_{2}-v_{1},P_{s}-K_{2}-v_{1})$ with $K_{2} \equiv K-v_{2}$. An eigenstate will be delocalized only along one of the lines on the $P_{s}$-plane. 
Two points are to be noted at this stage. Firstly, the area of the $P_{s}$-planes decrease with increasing $v_{3}$ and the length of the lines on a given plane decreases with increasing $v_{2}$. Consequently, unperturbed normal modes should be expected for high excitations in the $(v_{2},v_{3})$ modes. Secondly, the number of state space points $(v_{1},K_{2}-v_{1},P_{s}-K_{2}-v_{1})$ on a given line determines the quantum number $\mu$, an excitation index, 
related to the phase space area enclosed by the $156$-resonance island in the classical phase space. 

\begin{figure*}[t]
\begin{center}
\includegraphics[height=3.5in,width=6.5in]{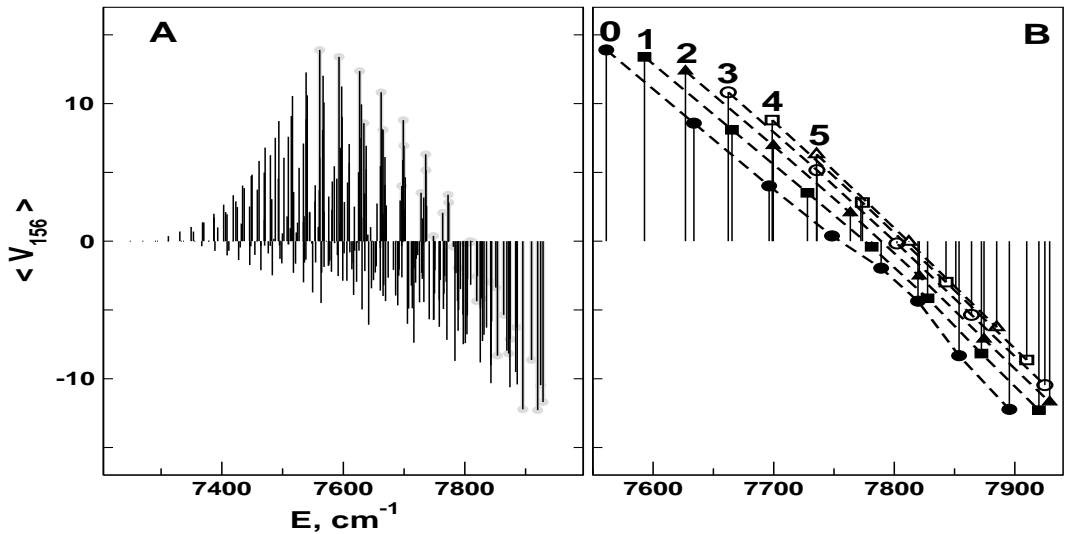}
\caption{Single $156$-resonance Hamiltonian. (A) Quantum expectation values $\langle V_{156} \rangle$ as a function of the eigenenergies.  The apparent complexity is due to the interspersing of several easily identifiable regular sequence of states. (B) $\langle V_{156} \rangle$ for states with $v_{3}=0$, {\it i.e.,} $P_{156} \equiv 2v_{1}+v_{5}+v_{6}=14$ (also shown in (A) as grey circles). 
Various sequences with differing $v_{2}$ are indicated by dashed lines. A sequence corresponding to say $v_{2}=0$ (filled circles) is assigned as $|7,14,0,14;0,\mu \rangle$ with $\mu=0,1,\ldots,7$ being an excitation index. See text for explanations.}
\label{lvelint156plot}
\end{center}
\end{figure*}

Therefore,  a knowledge of the quantum number $\mu$ provides a complete assignment $|K,L,M,P_{s};v_{2},\mu \rangle$ of the single resonance eigenstates. This information can be obtained both from the semiclassical wavefunction and the level-velocity approaches mentioned above. For instance, looking at the phase of the semiclassical state in the $\psi_{1}=0$ plane will immediately reveal the $(v_{2},v_{3})$ quantum numbers. Similarly, keeping track of the phase advance along the $\psi_{1}$ direction in the $\psi_{3}=0$ or $\psi_{2}=0$ plane will yield $\mu$ as the excitation quantum number\cite{jtjcp10,jtjpc07}. 

Equivalently, in Fig.~\ref{lvelint156plot}(A) the quantum eigenstate expectation values of the $156$-resonance operator are shown. The apparent congested nature of the figure is a direct result of the interleaving of the various state sequences with different $v_{3}$. The sequences, however, can be easily identified by analyzing the corresponding classical phase space function $v_{156}({\bf J};K_{c},L_{c}) \cos \psi_{1}$ with
\begin{equation}
v_{156} \equiv 
 2\left[J_{1}(K_{2c}-J_{1})(P_{sc}-K_{2c}-J_{1}\right]^{1/2}.
\end{equation}
In the above expression the quantities $K_{2c}$ and $P_{sc}$ correspond to the classical analogs of the quantum $K_{2}$ and $P_{s}$ respectively.
It can be shown that classical  expectation value is large and has equal magnitudes at the resonant angles $\bar{\psi}_{1}=0,\pi$ and resonant action
\begin{equation}
\bar{J}_{1}=\frac{1}{3}\left[P_{sc}-\sqrt{P_{sc}^{2}-3K_{2c}(P_{sc}-K_{2c})}\right].
\label{clvel156predict}
\end{equation}
The expectation values are positive for $\psi_{1}=0$ and negative for $\psi_{1}=\pi$. From the classical phase space perspective the opposite signs translate to states being localized about the stable or unstable fixed points of the resonance island. Using the classical estimate it is possible to identify the highly localized states which further spawn the sequences with different $\mu$ and  in Fig.~\ref{lvelint156plot}(B) some of the $v_{2}$-sequences for fixed $v_{3}=0$ {\it i.e.,} $P_{s}=14$ are shown. For instance, the sequence shown in Fig.~\ref{lvelint156plot}(B) (circles) is assigned as $|7,14,0,14;0,\mu \rangle$ with $\mu=0$ (maximum positive expectation value) to $\mu=7$ (maximum negative expectation value). All such sequences can be easily identified and hence a complete assignment of the integrable, single resonance cases can be provided in terms of the appropriate quantum numbers.

Including other independent resonances in eq.~\ref{qham} results in a non-integrable system with highly mixed states which are presumably interspersed with some of the regular states. For example, including the $526$-resonance (and/or the $125$-resonance) in addition to the $156$-resonance destroys the $v_{2}$ quantum number. In terms of the geometry described above, since $v_{3}$ is still conserved, the $P_{s}$-planes are  preserved but the states now spread over the entire plane. Further inclusion of the $231$-resonance and $36$-resonance destroys the $P_{s}$-planes as well and the states now  spread over the entire state space. In the next two sections the goodness of $v_{2}$ and $v_{3}$ are broken sequentially in order to check if any new state sequences appear despite the nonintegrability of the system.

\subsection{Breaking the goodness of $v_{2}$}

\begin{center}
\begin{figure}[tbp]
\includegraphics[height=3.5in,width=3.5in]{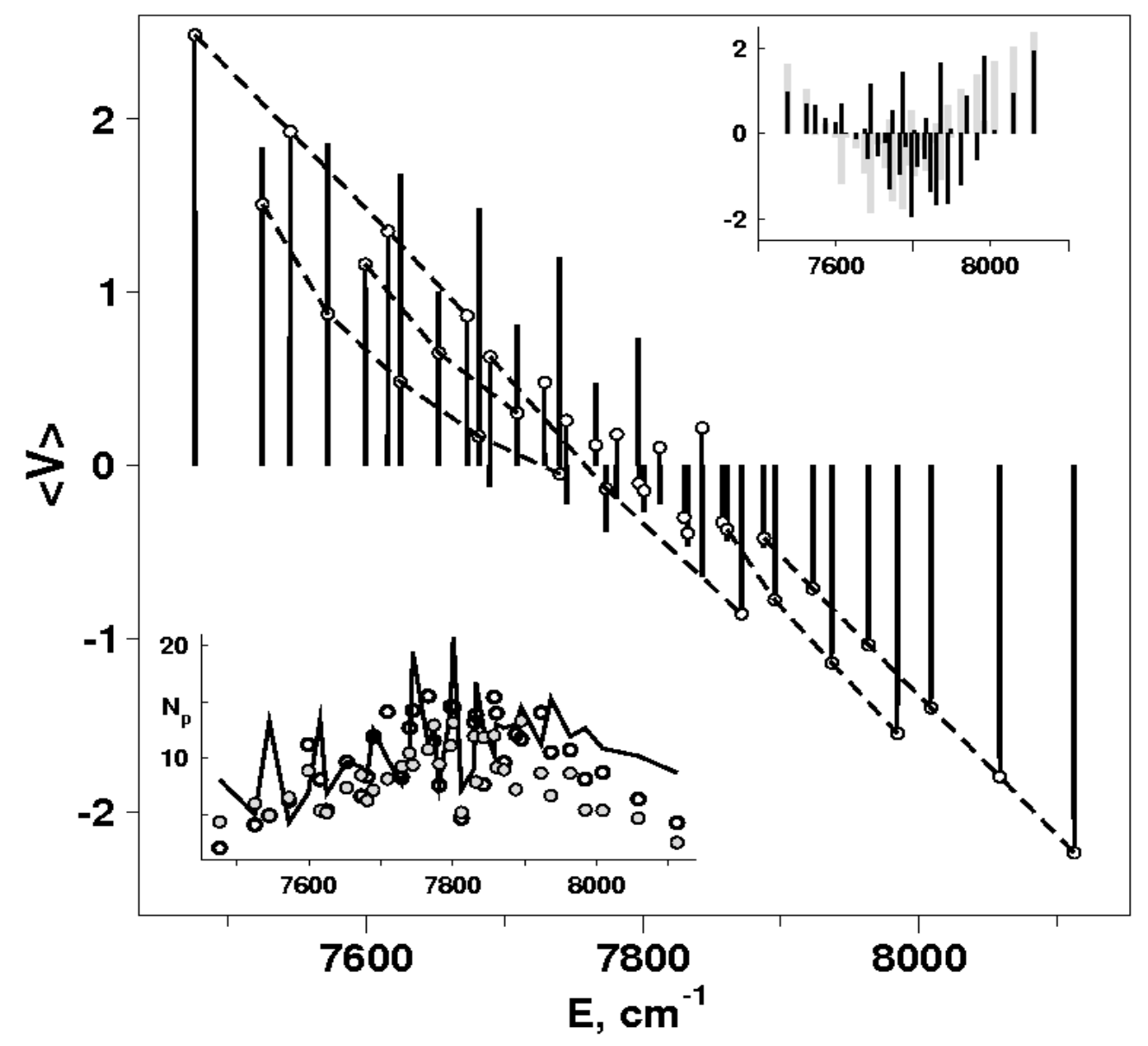}
\caption{Scaled level velocities for the noninetgrable system with $156$ (thick black lines), $526$ (thin lines with open circles), $125$ (black lines in top inset), and the $261$ (grey lines in top inset) anharmonic resonances present.  Eigenstates belonging to the good quantum number $v_{3}=0$ are shown. Two eigenstate sequences (dashed lines) starting at the high energy end are easily identified. Similar sequences at the low energy end are also shown. The sequences merge towards the middle resulting in mixed states. (Bottom inset) Participation ratios $N_{p}$ (cf. eq.~\ref{ipr}) in the $156$ (grey circles), $526$ (black open circles), and $125$ (black line) single resonance basis. }
\label{lvelpsgoodplot}
\end{figure}
\end{center}

In this section only the resonances involving modes $\nu_{1},\nu_{2},\nu_{5}$, and $\nu_{6}$ are considered. Thus, $v_{2}$ is no longer a good quantum number whereas $v_{3}$ is still a good quantum number. Note that the set of resonances involved here are the strongest ones in thiophosgene and therefore strong state mixings are expected. Are there any patterns in this case?  In analogy to Fig.~\ref{lvelint156plot}(B), the $v_{3}=0$ case of the nonintegrable system is shown in Fig.~\ref{lvelpsgoodplot} using the level velocity approach. Since several level velocities are compared in the figure, all of them are scaled to zero mean and unit variance.

From Fig.~\ref{lvelpsgoodplot} it is clear that eigenstate sequences do exist in this non-integrable case and a few of them are highlighted by dashed lines. It is important to note that the observed sequences are different from the $v_{2}$-sequences seen in Fig.~\ref{lvelint156plot}(B) since $v_{2}$ is no longer a good quantum number. Three key observations are worth making at this stage. Firstly, the initial sequences starting at high energy can be identified using $\langle V_{156} \rangle$ or equally well using $\langle V_{526} \rangle$, two of the strongest resonances in the system. Interestingly, as evident from Fig.~\ref{lvelpsgoodplot}(top inset), these sequences can also be identified from  $\langle V_{125} \rangle$ and $\langle V_{261} \rangle$ values. The large values of $\langle V_{261} \rangle$ despite the weakness of the $261$-resonance is an example of the nontrivial nature of such multiresonant systems. The two major resonances induce a relatively strong $261$-resonance in certain regions of the $P_{s}$ plane, resulting in significant perturbation of the state sequences. Thus, at high energies the states are under the influence of several resonances. Secondly, at low energies the sequences are less robust (note the curvature of the dashed lines in the figure)  and fewer in number, but seem to be organized more by the $526$-resonance. 
Thirdly, the various sequences  ``collide" near $E \sim 7800$ cm$^{-1}$ leading to significant disruption of the regularity of the sequences. Small level velocities and energy spacings in this region imply several multistate avoided crossings and suggest that the state mixing could be due to dynamical tunneling\cite{keirpc07}. Dynamical assignments of the states in this complicated transition region are difficult if not impossible.

Support for the level velocity predictions comes from computing the participation ratio $N_{P}$ in the various single resonance basis $|{\bf r} \rangle$ (eigenstates of single resonance Hamiltonians), shown in  Fig.~\ref{lvelpsgoodplot}(bottom inset).  The number of basis states participating in a given eigenstate $|\alpha \rangle$ is computed as
\begin{equation}
N_{P} = \left(\sum_{{\bf r}} |\langle {\bf r}|\alpha \rangle|^{4} \right)^{-1}.
\label{ipr}
\end{equation}
The participation ratios indicate that, except for a few states at the high and low energy ends, most states are moderately mixed. Moreover, large values of $N_{P}$ at $E \sim 7800$ cm$^{-1}$ in every basis agrees well with the level velocity data.
Note that hints for the existence of eigenstate sequences are present in the $N_{P}$ data, but nowhere as emphatically as reflected in the level velocity plots.

\begin{figure*}[t]
\begin{center}
\includegraphics[height=3.5in,width=5.5in]{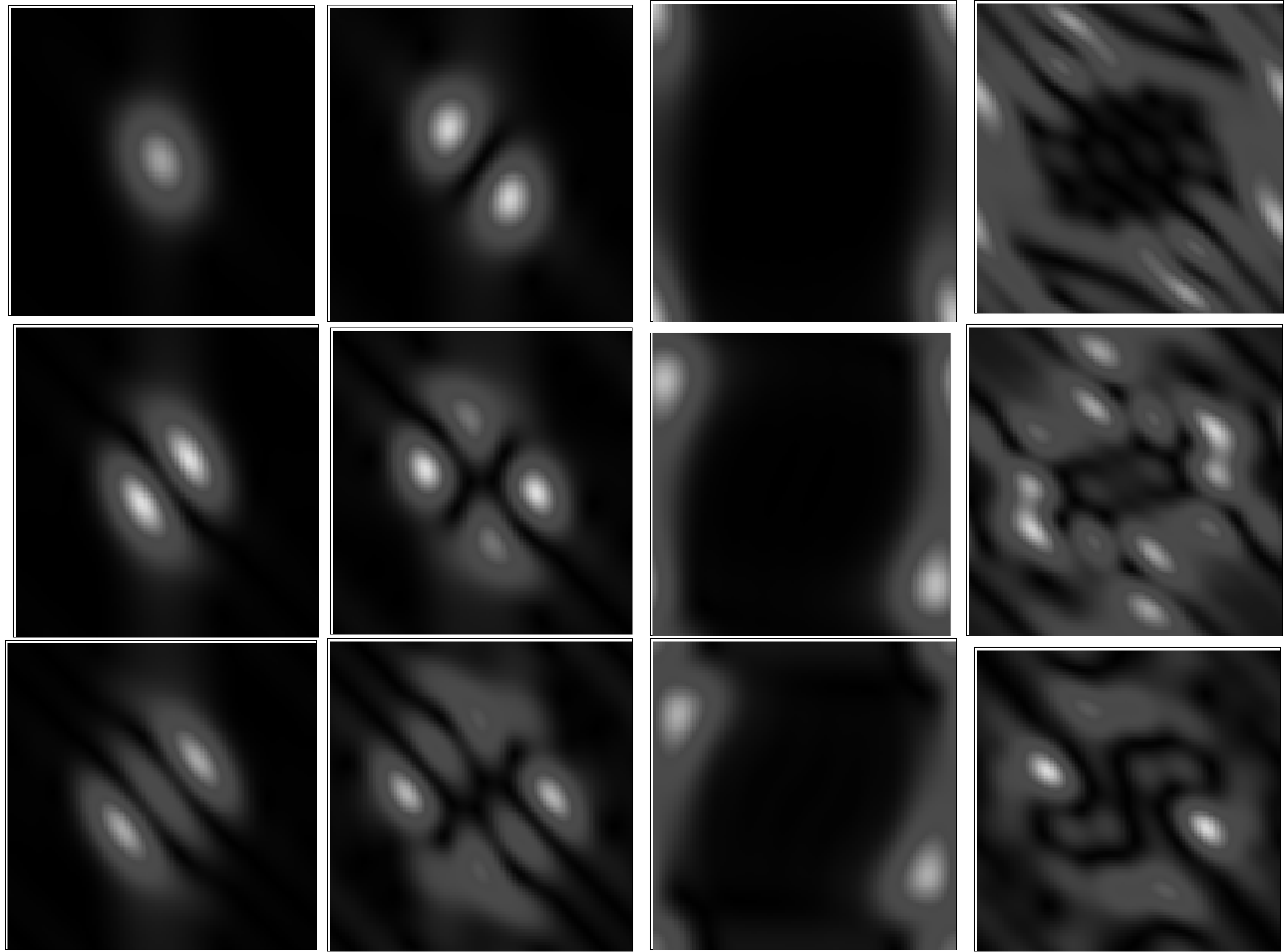}
\caption{Semiclassical $(\psi_{1},\psi_{2})$ angle space representations of select states corresponding to the case shown in Fig.~\ref{lvelpsgoodplot} with $v_{3}=0$ being a good quantum number. The $\psi_{3}=0$ slice of the full three dimensional space are shown. In every plot the angles range from $0$ to $2\pi$. (First column) States belonging to the first sequence in Fig.~\ref{lvelpsgoodplot} starting at the highest energy. (Second column) States forming the second sequence in Fig.~\ref{lvelpsgoodplot} at high energy. (Third column) States belonging to the first sequence in Fig.~\ref{lvelpsgoodplot} starting at the lowest energy. (Last column) States that do not form any clear sequence. See text for discussions.}
\label{psgoodsemi}
\end{center}
\end{figure*}

The various sequences observed here are consistent\cite{jtjcp10} with the previous results of Jung and Taylor. This can be established by looking at appropriate slices of the semiclassical three-dimensional angle space representations of the eiegnstates. In Fig.~\ref{psgoodsemi} the semiclassical $(\psi_{1},\psi_{2};\psi_{3}=0)$ angle space representations are shown for select states in the $v_{3}=0$ manifold. A striking correspondence can be seen between the sequences revealed by the level velocities shown in Fig.~\ref{lvelpsgoodplot} and the eigenstate density patterns as observed in the semiclassical angle space. The first column of states correspond to the sequence, indicated in Fig.~\ref{lvelpsgoodplot}, starting with the highest energy state and have clear nodal structure - no nodes in the diagonal direction and increasing number of nodes in the antidiagonal direction. Similarly, states in the second column belong to the second high energy sequence indicated in Fig.~\ref{lvelpsgoodplot} and exhibit an additional node in the diagonal direction. Both these sequences correspond to the class A classification given earlier\cite{jtjcp10} and the nodes along the diagonal and antidiagonal directions yield two additional quantum numbers $t_{d}$ and $t_{a}$ respectively. For instance, the first state in column one of Fig.~\ref{psgoodsemi} is assigned as $|K,L,M;v_{3}=0,t_{d}=0,t_{a}=0\rangle$
and the assignment $|K,L,M;0,1,2\rangle$ is appropriate for last state in the second column. 

On the other hand, the third column in Fig.~\ref{psgoodsemi} corresponds well to the sequence indicated in Fig.~\ref{lvelpsgoodplot} starting with the lowest energy state. Although the angle space densities show localization around $\psi_{1} = 0, 2\pi$, suggesting $156$-resonant states, the level velocity results show that the influence of the $526$-resonance is far greater in comparison. State space plots and the participation ratio data do establish the first state of the sequence to be closer to being a $526$-resonant state. However, the later states in the sequence are perturbed strongly by the induced $261$  resonance (see the top inset of Fig.~\ref{lvelpsgoodplot}) and hence nontrivial to assign. The last column of Fig.~\ref{psgoodsemi} show three states that are in the complicated energy region around $7800$ cm$^{-1}$. The mixed nature of these states is evident both from the angle space densities as well as the level velocity and participation ratio data. Presumably the induced $261$-resonance is leading to the highly mixed states. In particular, the middle state of the last column seems to be a state that is delocalized over an extended region of the phase space and such states have been observed in a different context in an earlier study\cite{mkjpc09}.

\begin{figure*}[htbp]
\begin{center}
\includegraphics[height=3.5in,width=6.5in]{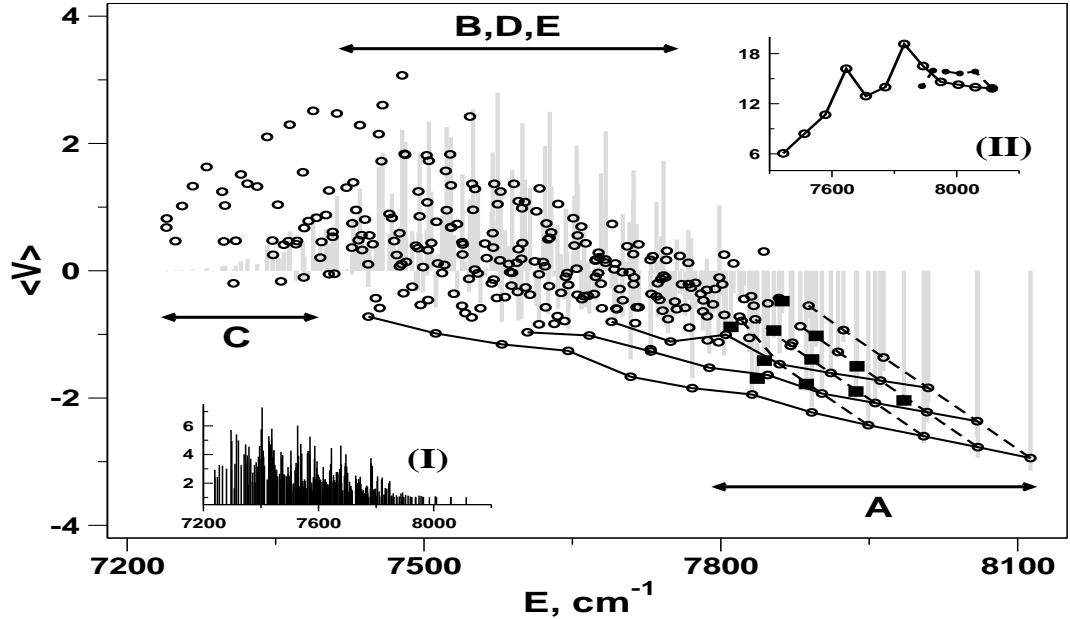}
\caption{Eigenstate sequences as revealed by $\langle V_{156} \rangle$ (in grey) and $\langle V_{526} \rangle$ (circles) forthe full Hamiltonian. The various classes of eigenstates are denoted by A,B,C,D, and E in accordance with an earlier study\cite{jtjcp10}. Sequences with increasing $v_{3}$ with fixed $(t_{d},t_{a})$ (black lines). Sequences with increasing $t_{d}$ with $t_{a}=0$ and fixed $v_{3}$ (dashed lines). States with varying $v_{3},t_{d}$ and fixed $t_{a}=1$ (squares). (I) Participation ratio of the eigenstates in the $v_{3}$-good basis. Note the significant breakdown of the goodness of $v_{3}$ in the middle of the polyad. (II) Participation ratio in the zeroth order basis of the $(t_{d},t_{a})=(0,0)$ sequence (solid line) and $v_{3}=0$ sequence (dashed lines).}
\label{fullplot}
\end{center}
\end{figure*}

\subsection{Breaking the goodness of $v_{3}$ - the full system}

The previous section identified sequences of eigenstates exhibiting similar localization properties of a subsystem of the full Hamiltonian eq.~\ref{qham} wherein $v_{3}$ was a good quantum number. The goodness of $v_{3}$ is strictly broken in the full system, obtained from the subsystem with the inclusion of the remaining $36$ and the $231$ resonances. A key question is whether the sequences in Fig.~\ref{lvelpsgoodplot} would persist in the full system. The answer to this question can be obtained by analyzing Fig.~\ref{fullplot} which shows scaled $\langle V_{156} \rangle$ and $\langle V_{526} \rangle$ for all the $288$ eigenstates. The expectation values $\langle V_{261} \rangle$ and $\langle V_{125} \rangle$ are also substantial, as in Fig.~\ref{lvelpsgoodplot}, but not shown  for the purpose of clarity. It is immediately clear from Fig.~\ref{fullplot} that the high end of the polyad does exhibit several sequences highlighted by solid and dashed lines connecting the states. Specifically, the solid line sequences correspond to increasing $v_{3}$ with fixed $(t_{d},t_{a})$ quantum numbers and the dashed lines correspond to increasing $t_{d}$ with $t_{a}=0$ and fixed $v_{3}$ values. In particular, these high energy states correspond precisely to the class A states discussed in the previous section with similar semiclassical $(\psi_{1},\psi_{2})$ angle space representations. Note that in Fig.~\ref{lvelpsgoodplot} only the $v_{3}=0$ states are shown and including states for different values of $v_{3}$ would show high energy patterns similar to those seen in Fig.~\ref{fullplot}. 

The robustness of the class A sequences is also confirmed in Fig.~\ref{fullplot}(I) wherein the participation ratio of the high energy eigenstates is seen to be nearly unity in the $v_{3}$-good basis {\it i.e.,} in the eigenbasis of the subsytem discussed in the previous section. For these set of states $v_{3}$ is approximately conserved and can be used to label the eigenstates. Consequently, it is possible to infer that the lowest frequency mode $\nu_{3}$ ($a_{1}$, Cl-C-Cl bend) is decoupled from the rest of the modes in the $[7800,8100]$ cm$^{-1}$ region. The IVR dynamics ensuing from a feature (anharmonic zeroth order) state comprising mainly of the class A states is therefore expected to be nonstatistical. A crucial observation in this context is summarized in Fig.~\ref{fullplot}(II) - participation ratios in the zeroth-order basis are unable to identify the various sequences as reflected in the expectation value patterns. Stated differently, delocalization in state space does not immediately imply delocalization in the semiclassical angle space and hence in the full phase space.

Furthermore, two important observations  can be made in the mid to low energy regions of the polyad.
Firstly,  for $E \lesssim 7800$ cm$^{-1}$ it is clear from Fig.~\ref{fullplot}(I) that the goodness of $v_{3}$ is broken substantially.  Indeed, the regular lattice formed by the various sequences at the high energy end of the polyad begins to ``dissolve" around $E \sim 7800$ cm$^{-1}$ due to the influence of several resonances. Secondly, at the lowest energy end, the states are mainly influenced by the $526$-resonance and hence, as noted earlier, typical class C states. The middle region of the polyad spanning $[7350,7750]$ cm$^{-1}$ is particularly complicated with several highly mixed states and states exhibiting transitional nature between the two extreme classes, A and C, of localized states. 

\section{Highly excited $\nu_{4}$ states}

Up until now the focus has been exclusively on the $v_{4}=0$ case {\it i.e.,} no excitations in the out-of-plane bending mode. Experimentally, however, the $\nu_{1}$ and $\nu_{4}$ modes have dominant Franck-Condon activity\cite{sgpccp04} and sharp assignable features in the spectra are seen\cite{cgjcp109} even around and above the dissociation energy. Most of the sharp features have fairly large values of $v_{4}$ in their assignments. Thus, it is interesting to ask if the nature of the eigenstates, in terms of localization as well as the different classes, undergo any significant change upon exciting the $\nu_{4}$ mode. Earlier works have hinted at a smooth morphing of the reduced dimensional wavefunctions\cite{cgjcp109} with increasing $\nu_{4}$ excitation and the existence of IVR-protected zeroth-order states for large number of quanta in the $\nu_{4}$ mode\cite{sgjcp06}. Can the level-velocity technique used in this work provide any insights? 

\begin{figure}[h]
\includegraphics[height=3.5in,width=3.5in]{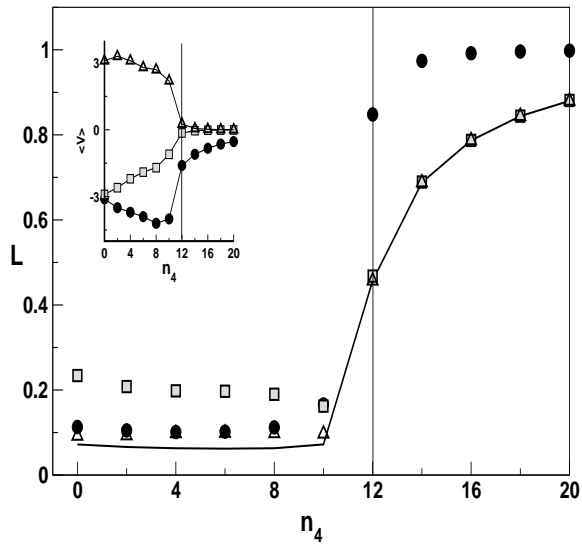}
\caption{Variation of the participation ratios of the highest energy eigenstate in the polyad $[7,14,v_{4}]$ with increasing excitation of the out-of-plane bending mode $\nu_{4}$. The zeroth-order basis, $156$, $526$, and $125$ single resonance basis values are shown as black line, filled circles, grey squares, and triangles respectively. Inset shows the expectation values $\langle V_{156} \rangle$ (filled circles), $\langle V_{526} \rangle$ (grey squares), and $\langle V_{125} \rangle$ (triangles) for the highest energy state as a function of $v_{4}$. Note the sharp transition in both measures around $v_{4}=12$.}
\label{n4bifurplot}
\end{figure}

As a preliminary effort, Fig.~\ref{n4bifurplot} summarizes the result of increasing $\nu_{4}$ excitation on the highest energy eigenstate in the polyads $[K=7,L=14,v_{4}]$. The total number of eigenstates remains the same ($288$) due to the decoupling of the $\nu_{4}$ mode from the other modes, but the eigenstates span higher energy ranges. The main part of Fig.~\ref{n4bifurplot} shows the participation ratios of the eigenstate in various basis. Clearly, the highest energy state changes it nature from the one shown in Fig.~\ref{psgoodsemi} (first state of the first column) to one that is highly localized. Looking at the state space representation of the eigenstate for $v_{4}=20$ it becomes clear that it is a pure normal mode state. More importantly, the participation ratios undergo a sharp transition around $v_{4}=12$, signaling a significant change in the nature of the eigenstate. It is intriguing to observe that the $156$-resonance basis value exhibits a much sharper transition as compared to the other basis. 

The various scaled expectation values shown in the inset of Fig.~\ref{n4bifurplot} confirm the results based on the participation ratios. In particular, an equally sharp transtition is also seen in this case at $v_{4}=12$ and hence establishes the sensitivity of the level-velocities to the changing nature of the eigenstate. For $v_{4} \geq 12$ all the expectation values are close to zero and thus predict a pure normal mode state. Incidentally, the results here agree well with the observation of IVR-protected zeroth-order states around $v_{4}=10$ in an earlier study\cite{sgjcp06} by Sibert and Gruebele (see figure $7$ of their paper). However, it is interesting to note that the expectation values take on maximum values where the participation ratios are minimum. Similar observations\cite{skcpl04} were made in an earlier study regarding the appearance of new modes in highly excited acetylene. Thus, the results of Fig.~\ref{n4bifurplot} suggest that there is a bifurcation in the underlying classical phase space which could be responsible for the experimental observation of sharp features near the dissociation threshold. Undoubtedly, more work is needed to establish the detailed bifurcation structure of the classical phase space and corresponding fingerprints on the quantum eigenstates. Perhaps, such an  analysis would resolve the puzzle as to why such sharp features can be predicted\cite{cgjcp109} with an effective Hamiltonian without any resonance couplings.

\section{Conclusions}

The central point of this work is to emphasize the fact that highly excited eigenstates of multiresonant effective Hamiltonians form several sequences, corresponding to different classes of dynamics,  interspersed amongst each other in a complicated fashion. However, the tools used in this work are ideal to disentangle the sequences and gain insights into the nature of the eigenstates. A remarkable aspect that emerges from this and other previous studies is worth noting - the expectation values shown, for example, in Fig.~\ref{fullplot} are quantum objects. Nevertheless, the observed patterns and sequences in the expectation values have a one-to-one correspondence with the semiclassical angle space densities of the eigenstates. Thus, this establishes the close classical-quantum correspondence that exists for spectroscopic Hamiltonians.  At the same time, the observation that state space based measures of localization or delocalization of the eigenstates are insufficient to identify the sequences suggests that phase space is the ideal setting where the true nature of the eigenstates are revealed - a conclusion that fortifies decades of earlier work\cite{dynassign} on systems with smaller degrees of freedom. The semiclassical action-angle representation is not all that abstract afterall!

It is heartening to see that the simple level-velocity approach is capable of identifying the eigenstate sequences. The present work agrees, to a large extent, with the previous analysis\cite{jtjcp10} and classification of the eigenstates. In particular, the eigenstate classes predicted in the earlier studies do persist in this study which uses a more accurate Hamiltonian. Therefore, the different classes of eigenstates and predictions of their influence on the IVR dynamics of certain zeroth-order bright states are expected to be robust even upon inclusion of other weak higher order resonances. The approach adopted in this work is easily generalized to systems with any number of degrees of freedom  with several anharmonic resonances. The recently proposed method\cite{mkjpc09} of lifting the eigenstates onto the Arnold web {\it i.e.,} the network of nonlinear resonances can lead to further detailed insights into the anatomy of certain mixed states (cf. last column of Fig.~\ref{psgoodsemi}).

Finally, Fig.~\ref{psgoodsemi} presents another crucial question. Are there experimental signatures of the eigenstate sequences? Clearly, a state localized in the semiclassical angle space representation need not be localized in the state space representation. However, since the quantum expectation values do pick out the sequences correctly, it is of some interest to see if there are any spectral signatures of such different classes of eigenstate sequences. Alternatively, Fig.~\ref{n4bifurplot} suggests that new modes, like the counter-rotator and local bending modes in acetylene, might appear in SCCl$_{2}$ due to the various bifurcations in the underlying classical phase space. Again, there must be spectral fingerprints of such new modes. More work in this direction needs to be done.

\section{Acknowledgments}
It is a pleasure to thank Martin Gruebele and Ned Sibert for sending the full set of parameters of the fourth order effective Hamiltonian for SCCl$_{2}$.

\pagebreak


\begin{thebibliography}{99}
\bibitem{smalley83}{See for example,
Lehmann, K. K.; Scoles, G.; Pate, B. H.
{\it Ann. Rev. Phys. Chem.} {\bf 1994}, {\it 45}, 241;
Nesbitt, D. J.; Field, R. W. {\it J. Phys. Chem.}
{\bf 1996}, {\it 100}, 12735;
Gruebele, M; Bigwood, R. {\it Int. Rev. Phys. Chem.}
{\bf 1998}, {\it 17}, 91;
Keske, J. C.; Pate B. H. {\it Annu. Rev. Phys. Chem.}
{\bf 2000}, {\it 51}, 323.}
\bibitem{bnn04}{B\"{a}ck, A.; Nordholm, S.; Nyman, G. {\it J. Phys. Chem. A}
{\bf 2004}, {\it 108}, 8782.}
\bibitem{jajcp99}{Jacobson, M. P.; Jung, C.; Taylor, H. S.; Field, R. W.
{\it J. Chem. Phys.} {\bf 1999}, {\it 111}, 600; Jung, C.; Mejia-Monasterio, C.; Taylor, H. S.
{\it J. Chem. Phys.} {\bf 2004}, {\it 120}, 4194.}
\bibitem{gl08}{Leitner, D. M.; Gruebele, M. {\it Mol. Phys.} {\bf 2008},
{\it 106}, 433.}
\bibitem{dynassign}{Kellman, M. E.; Tyng, V. {\it Acc. Chem. Res.} {\bf 2007},
{\it 40}, 243. Ezra, G. S. {\it Adv. Clas. Traj. Meth.} {\bf 1998},
{\it 3}, 35. Davis, M. J. {\it Int. Rev. Phys. Chem.} {\bf 1995},
{\it 14}, 15. Farantos, S. C.; Schinke, R.; Guo, H.; Joyeux, M. {\it Chem. Rev.} {\bf 2009}, {\it 109}, 4248.}
\bibitem{sgjcp06}{Sibert, E. L.; Gruebele, M. {\it J. Chem. Phys.} {\bf 2006}, {\it 124}, 024317.}
\bibitem{sgpccp04}{Strickler, B.; Gruebele, M. {\it Phys. Chem. Chem. Phys.} {\bf 2004}, {\it 6}, 3786.}
\bibitem{cgjcp109}{Chowdary, P. D.; Gruebele, M. {\it J. Chem. Phys.} {\bf 2009}, {\it 130}, 024305.}
\bibitem{rmjcp08}{Rashev, S.; Moule, D. C. {\it J. Chem. Phys.} {\bf 2008}, {\it 128}, 091101.}
\bibitem{cgjcp209}{Chowdary, P. D.; Gruebele, M. {\it J. Chem. Phys.} {\bf 2009}, {\it 130}, 134310.}
\bibitem{jtsjpc06}{Jung, C.; Taylor, H. S.; Sibert, E. L. {\it J. Phys. Chem. A} {\bf 2006}, {\it 110}, 5317.}
\bibitem{jtjcp10}{Jung, C.; Taylor, H. S. {\it J. Chem. Phys.} {\bf 2010}, {\it 132}, 234303.}
\bibitem{jtjpc07}{Jung, C.; Taylor, H. S. {\it J. Phys. Chem. A} {\bf 2007}, {\it 111}, 3047.}
\bibitem{sejcp03}{Keshavamurthy, S. {\it J. Phys. Chem. A} {\bf 2001},
{\it 105}, 2668.
Semparithi, A.; Charulatha, V.; Keshavamurthy, S.
{\it J. Chem. Phys.} {\bf 2003}, {\it 118}, 1146.
Semparithi, A.; Keshavamurthy, S. {\it Phys. Chem.
Chem. Phys.} {\bf 2003}, {\it 5}, 5051.}
\bibitem{skcpl04}{Semparithi, A.; Keshavamurthy, S. {\it Chem. Phys. Lett.}
{\bf 2004}, {\it 395}, 327.}
\bibitem{mkjpc09}{Manikandan, P.; Semparithi, A.; Keshavamurthy, S. {\it J. Phys. Chem. A} {\bf 2009}, {\it 113}, 1717.}
\bibitem{peres}{Peres, A. {\it Phys. Rev. A} {\bf 1984}, {\it 30}, 504.
Eckhardt, B.; Fishman, S.; Keating, J.; Agam, O.;
Main, J.; M\"{u}ller, K. {\it Phys. Rev. E} {\bf 1995}, {\it 52}, 5893.
Mehlig, B.; M\"{u}ller, K.; Eckhardt, B.
{\it Phys. Rev. E} {\bf 1999}, {\it 59}, 5272. Boos\'{e}, D.; Main, J. {\it Phys. Rev. E} {\bf 1999},
{\it 60}, 2831.}
\bibitem{gjpc96}{Gruebele, M. {\it J. Phys. Chem.} {\bf 1996}, {\it 100}, 12178.}
\bibitem{keirpc07}{Keshavamurthy, S. {\it Int. Rev. Phys. Chem.} {\bf 2007},
{\it 26}, 521.}


\end{thebibliography}
\end{document}